\documentclass[
 reprint,
 amsmath,amssymb,
 aps,
 pra,
 superscriptaddress]{revtex4-2}

\usepackage{graphicx}% Include figure files
\usepackage{dcolumn}% Align table columns on decimal point
\usepackage{bm}% bold math

\usepackage{xcolor}

\begin{document}

\preprint{APS/123-QED}

\title{Dynamics of discrete spacetimes with Quantum-enhanced Markov Chain Monte Carlo}

\author{Stuart Ferguson}
\email{S.A.ferguson-3@ed.ac.uk}
\affiliation{%
 Quantum Software Lab, School of Informatics, The University of Edinburgh, Edinburgh, United Kingdom
}%

\author{Arad Nasiri}
\affiliation{Blackett Laboratory, Imperial College London, London, United Kingdom}
\author{Petros Wallden}
\affiliation{%
 Quantum Software Lab, School of Informatics, The University of Edinburgh, Edinburgh, United Kingdom
}%

\date{\today}

\begin{abstract}

Quantum algorithms offer the potential for significant computational advantages; however, in many cases, it remains unclear how these advantages can be practically realized. Causal Set Theory is a discrete, Lorentz-invariant approach to quantum gravity which may be well positioned to benefit from quantum computing. In this work, we introduce a quantum algorithm that investigates the dynamics of causal sets by sampling the space of causal sets, improving on classical methods. Our approach builds on the quantum-enhanced Markov chain Monte Carlo technique developed by Layden et al. [Nature 619, 282 (2023)], adapting it to sample from the constrained spaces required for application. This is done by adding a constraint term to the Hamiltonian of the system. A qubit Hamiltonian representing the Benincasa-Dowker action (the causal set equivalent of the Einstein-Hilbert action) is also derived and used in the algorithm as the problem Hamiltonian. We achieve a super-quadratic quantum scaling advantage and, under some conditions, demonstrate a greater potential compared to classical approaches than previously observed in unconstrained QeMCMC implementations.

\end{abstract}

\maketitle

\textit{Introduction}\textemdash Causal Set Theory (CST) is one approach to quantum gravity that assumes spacetime to be fundamentally discrete, without violation of local Lorentz invariance. First developed by Bombelli et al. \cite{bombelli1987}, one of its early successes was the correct prediction of the order of magnitude of the cosmological constant \cite{sorkin1991spacetime}, followed by phenomenological models that implemented and tested the idea against cosmological data \cite{Ahmed:2002mj,Zwane:2017xbg,Das:2023hbw,Das:2023rvg}. On the theoretical front, sampling causal sets according to a partition function is instrumental in understanding the dynamics of causal sets. This is typically addressed computationally utilizing Markov Chain Monte Carlo (MCMC) \cite{surya2012evidence, henson2015, glaser2016, glaser2018finite, glaser2021, Glaser_2018, cunningham2018, cunningham2020}. In the studied cases, the computational complexity is exponential, limiting analysis to small causal sets, representing tiny portions of spacetime. Here, we introduce a set of novel numerical methods to sample the space of causal sets of fixed cardinality. We build on the Quantum-enhanced MCMC (QeMCMC) of Layden et al. \cite{layden2023}, by adapting the algorithm to constrained problems and providing an example in sampling causal sets.

Our primary contributions are twofold: (i) the design of a quantum algorithm to sample causal sets uniformly, and (ii) the development of a second algorithm to sample according to the Benincasa–Dowker (BD) action, which is the causal sets equivalent to the Einstein-Hilbert action. Both algorithms require restricting the sample space from general directed acyclic graphs to the subspace of valid causal sets. To enforce this constraint, we introduce and analyze the use of Hamiltonian constraints within the QeMCMC framework. For the second algorithm, we construct a qubit Hamiltonian that encodes an approximation of the BD action, enabling weighted sampling under the same constrained QeMCMC process. These components represent two key technical contributions: a constrained QeMCMC and a BD-action qubit Hamiltonian.

\textit{Causal set theory}\textemdash CST posits causal relations between a set of discrete spacetime elements as the fundamental degrees of freedom \cite{bombelli1987,surya2019causal}. These discrete spacetime elements form partially ordered sets whose adjacency matrix, also known as the causal matrix $C_{xy}$, encodes whether an element $x$ is in the causal past of $y$, denoted $x\prec y$. Importantly, the causal matrix should be transitive. Mathematically,  ($\forall x, y, z \in C)(x \prec y,\ y \prec z \Longrightarrow x \prec z$). We work with natural labellings of the causal sets, where the labels $x$ for the elements can be taken to be natural numbers such that if $x\prec y$ then $x<y$. %\arad{do we need extra operations after MCMC moves to bring them to naturally labelled form?}

Since explicit general covariance is central to CST, studies on quantum dynamics of causal sets have mostly focused on the sum-over-histories quantization in the form of path sums over causal sets. The partition function of causal sets of size $N$ is given by
\begin{align}
\label{eqn:Z}
    Z[N,\varepsilon]=\sum_{|C|=N}\mu(C)e^{iS_\varepsilon^{(d)}[C]}.
\end{align}
Here, $\mu(C)$ is a measure factor that would ultimately be fixed given a decoherence functional over causal sets, by integrating out the backward path in Schwinger-Keldysh formalism; see Appendix \ref{appendix_from_D_to_Z}. %A given choice of the decoherence functional determines the histories Hilbert space of causal sets, as well as their dynamics \cite{dowker2010hilbert}. In particular, it explicitly encodes what causal sets are allowed to interfere. 
As mentioned in \cite{cunningham2020}, it could be that $\mu(C)$ is not a uniform measure, for instance one might impose a stronger weight on manifoldlike causal sets. For the current work, in the absence of any agreed-upon decoherence functional for causal sets, we don't make any assumptions about $\mu(C)$.

The above amplitude contains the Benincasa-Dowker (BD) action $S_\varepsilon$, which is a one-parameter family of actions for causal sets \cite{benincasa2010,dowker2013causal}. In 4d:

\begin{equation}
\begin{aligned}
\label{eqn:smeared_BD_action}
    S_{\varepsilon}^{(4)} & =\frac{4}{\sqrt6} \sqrt\varepsilon\left[N- \varepsilon \sum_{j=0}^{N-2}  f_4(j, \varepsilon)N_j\right] \\
f_4(j, \varepsilon) & =(1-\varepsilon)^j\Bigg[1-\frac{9 \varepsilon j}{1-\varepsilon}+\\ & \frac{8\varepsilon^2 j(j-1)}{(1-\varepsilon)^2}-\frac{4\varepsilon^3j!}{3(j-3)!(1-\varepsilon)^3}\Bigg]\,.
\end{aligned}
\end{equation}
This is a linear combination of the abundances $N_k$, counting pairs $x\prec y$ that have $k$ elements in between (excluding $x$ and $y$). After a suitable averaging over Poisson-sprinkled causal sets on a manifold and taking the continuum limit, this action reduces to the Einstein-Hilbert action. The fluctuations around the Einstein-Hilbert action are controlled by $\varepsilon$. The $\varepsilon\ll1$ case has the smallest fluctuations, while $\varepsilon=1$ has the simplest expression, which is only in terms of $N_1,...,N_4$. In what follows, we use the 4d action for simplicity and concreteness. Nevertheless, our results can be generalized to any dimension; see Appendix \ref{appendix_arbitrary_d}.

The first MCMC sampling over causal sets was conducted in \cite{surya2012evidence} over the restricted class of 2d orders. The restriction allowed for a natural MCMC step, without needing to check transitivity. This instigated a plethora of interesting investigations into sampling causal sets, many of which restricted the class of causal sets to simplify the sampling procedure \cite{glaser2016hartle, glaser2018finite, cunningham2018, cunningham2020}. It should be noted that to enable MCMC sampling, one must introduce an auxiliary parameter $\beta$ which is then analytically continued,  $\beta\rightarrow i\beta$. Thus all of these algorithms sample in the \textit{Euclidean} setting, not \textit{Lorentzian}. In \cite{henson2015}, an alternative MCMC sampling strategy was employed to find the onset of KR domination with a uniform measure over causal sets. The MCMC steps dealt with the transitivity problem head-on, performing extra operations to ensure transitivity. As expected in any algorithm approaching this problem, thermalization times scaled exponentially with cardinality of the causal set. %, limiting results up to $N=85$.

Although restricted in cardinality, for the problem of sampling from the complete space of causal sets, this is the state of the art classical algorithm that both quantum algorithms detailed in this work will be benchmarked against. To our knowledge, there currently exists no algorithm that attempts to sample the space of all causal sets weighted by BD action. One can easily alter the algorithm in \cite{henson2006}, however to our knowledge this has not been investigated until this work. Here, we also propose a quantum algorithm designed explicitly for this problem by deriving a Hamiltonian term that approximates BD action.

\textit{Quantum Algorithm}\textemdash The classical intractability of large cardinality MCMC within CST leads one to consider quantum algorithms for such problems. Recently, a Quantum-enhanced Markov Chain Monte Carlo (QeMCMC) algorithm has been introduced, with a quantum scaling advantage in the problem of sampling from the Boltzmann distribution of a fully connected Ising model \cite{layden2023, ferguson2025}. 

QeMCMC is a novel type of hybrid algorithm that closely resembles its classical counterpart, with the replacement of one subroutine. Traditionally, a simple spin flip is used to ``propose'' new states for the Markov chain to explore. In the quantum method, the system is first initialized in the basis state corresponding to the current configuration of the chain, %starting with the current configuration of the chain, 
before undergoing quantum real-time evolution \cite{suzuki1990} for a time $t$ under a specially designed Hamiltonian

\begin{equation}
\label{eqn:H_prob_H_mix}
    H = (1-\gamma)H_{prob} + \gamma H_{mix}\,.
\end{equation}

This process explores configurations of \textit{similar energy}, yet \textit{far in Hamming distance}--the optimal proposal features for ``glassy'' cost functions. Here, $H_{prob}$ is a Hamiltonian formulation of the problem cost function, and $H_{mix} = \sum_iX_i$ is a mixing term designed to perturb the system into a superposition of many different configurations. The strength of this mixing is governed by the parameter $\gamma$. When the system is measured in the computational basis, the resulting configuration is used as the proposal, leading to very fast thermalization when compared to classical methods for which the proposal is less well informed. The rest of the MCMC procedure is unchanged, which means that even in the presence of quantum noise, there is no induced error in the output of the algorithm. The result is to merely slow down the thermalization, returning to the classical thermalization in the limit of fully depolarizing noise. This uncharacteristic resilience to quantum noise is of particular importance when considering the incoming early fault-tolerant era of quantum computing.

\textit{Constrained space}\textemdash 
To adapt QeMCMC to a broader class of applications, including here in CST, we introduce a Hamiltonian term, $H_{constr}$ to constrain the sample space:
\begin{equation}
\label{eqn:H_prob_H_constrained}
    H = \gamma_{prob}H_{prob} + \gamma_{mix} H_{mix} + \gamma_{constr} H_{constr}\,.
\end{equation}
This term must be designed so that it is large for configurations that are outside the constraint space, while being zero for ``valid'' configurations. Assuming the quantum circuit is initialized in one of the degenerate ground states of $H_{constr}$, proposal to invalid excited states is suppressed. Of course, the probability of invalid proposals is a function of $\gamma_{mix}$ and $\gamma_{constr}$, as well as the magnitude of $H_{constr}$ for invalid configurations. Given this simple Hamiltonian, all that a practitioner must do is derive a $H_{constr}$ that constrains their sample space.

\textit{Quantum uniform sampling}\textemdash 
To adapt the QeMCMC to CST, we must first decide how to represent the causal set into qubits. We encode the $q = \frac{N(N-1)}{2}$ possible causal relations in $q$ bits, the resulting bitstring represents the upper triangle of a naturally labeled causal matrix, $C$ - the most instinctive representation of a causal set on a computer. There are $2^q \approx 2^{\frac{N^2}{2}}$ possible configurations, while there are only $\approx 2^{\frac{N^2}{4}}$ causal sets \cite{kleitman1975asymptotic}, thus if one were to sample the entire configuration space, they would effectively be sampling the space of directed acyclic graphs \footnote{A method for sampling causal sets has been developed that samples the space of such graphs, before performing transitive reduction. It may have very quick thermalization for small instances, but is reported to slow down at approximately $N \approx 50$. \citetext{private communication, S. Surya, 2025.}}.

To use a quantum computer to propose new configurations, the quantum state (before measurement) must be restricted to a superposition of configurations that are causal sets. Each of the $q$ relations are encoded directly onto $q$ qubits. In other words, each relevant element in the causal matrix is represented by a single qubit, so the state of that qubit represents $C_{ij}$, the causal relationship between $i$ and $j$. %\arad{Aren't the last two sentences repeating the point of the previous paragraph?}\stuart{Kind of? The last paragraph was discussing a mapping to \textit{bits}, not \textit{qubits}. I agree that it sounds repetitive but i wanted to make it clear as its actually not obvious. Let me think on it}

Initializing the quantum computer in a bitstring representing a known causal matrix, inspired by QeMCMC, we can perform time evolution under a mixing Hamiltonian which creates a controlled superposition of each state. In doing this, we accidentally superpose computational basis states that do not correspond to causal matrices. Thus, we must first add a Hamiltonian constraint that effectively applies Transitive Closure (TC), to ensure that bitstrings representing non-causal sets are suppressed in our superposition. For example, given all cardinality 3 directed acyclic graphs the only possible configuration that breaks transitivity is the one for which there is a relation between events $1$ and $2$, $2$ and $3$ but not $1$ and $3$. In other words, this is a causal set iff $C_{12}C_{23}(1-C_{13}) = 0$. Generalizing this for larger cardinality by summing over all sets of 3 elements, we can write a Hamiltonian term that adds penalty $rP$ (where $r$ is the number of triplets that break transitivity, and P is the penalty term) for all configurations that are not causal sets:
\begin{equation}
\label{eqn:H_TC}
    H_{TC} = P\sum_{i<j<k}^N C_{ij}C_{jk}(1-C_{ik})\,.
\end{equation}
Thus, all causal sets of a given cardinality are degenerate ground states of this Hamiltonian. 

To uniformly sample from causal sets with a quantum computer, we start with an arbitrary causal set. We then perform time evolution under a Hamiltonian $H_{uniform}$ made up from this $H_{TC}$ constraint and a mixing term:
\begin{equation}
\label{eqn:H_TC_H_mix}
\begin{aligned}
    H_{uniform} &= (1-\gamma)H_{constr} + \gamma H_{mix}\,\\
    &= (1-\gamma)H_{TC} + \gamma H_{mix}\,.
    %\\
    %&= (1-\gamma)\sum_{i<j<k}^N C_{ij}C_{jk}(1-C_{ik}) + \gamma \sum_i X_i
\end{aligned}
\end{equation}
Note that here there is no ``problem'' term, as the aim is to uniformly sample. To get a uniform distribution over causal sets, however, the time $t$ must be long and thus the probability of transition to the exponentially large set of ``forbidden'' non-causal set configurations would dominate. The solution is to use MCMC, where each proposal is generated by measuring the quantum state prepared by time evolution of the previous configuration under $H_{uniform}$. The Markov chain begins with a known causal set and updates by proposing a new configuration via this quantum method, accepting it only if it forms a valid causal set (checked classically). In this way, we can uniformly sample, without worrying about any exponential costs in $\gamma$ or $t$. To avoid having to continually adjust these parameters for different cardinality, each term must be normalized - for details, see Appendix \ref{app:alphas}.

%In this way, the evolution is restricted to shorter times which limits the probability of proposal to forbidden states. 
By the design of $H_{mix}$, this process satisfies ergodicity within the space of causal sets, as each causal matrix can be reached by a finite number of bit flips from every other causal matrix. Of course, this algorithm is deliberately not ergodic in the space of all binary upper triangular matrices (directed acylcic graphs). Detailed balance is satisfied as the probability of proposal from one causal set to another and the inverse of this process are equal. This is clear as they are functions of only Hamming distance and absolute relative energy - which are trivially equal in both directions.

\textit{Weighted sampling}\textemdash Sampling algorithms in CST don't always aim to explore the space of all causal sets uniformly, instead the partition function \eqref{eqn:Z} is explored in order to ascertain the dynamics. Algorithms currently described in the literature reduce the size of the space by restricting it in some way. This has been done by focusing on 2d orders \cite{surya2012evidence, glaser2021, Glaser_2018}, or by fixing the background spacetime \cite{cunningham2018, cunningham2020}.

Here, we introduce a quantum algorithm that can explore the unrestricted space. The QeMCMC subroutine is again pivotal, however we advance the aforementioned quantum uniform sampling algorithm by employing the Metropolis-Hastings acceptance criteria - as done in most classical algorithms: a move from a causal set $C$ to $C'$ is accepted if $\exp \left(-\beta \Delta S_{\epsilon}^{(d)}\right)>u$, where $u$ is a uniform random variable in the range $[0,1)$ \cite{surya2012evidence, cunningham2020, metropolis1953}. This is for the case when the weights $\mu(C)$ in Equation~\eqref{eqn:Z} are uniform; otherwise, refer to the end of Appendix~\ref{appendix_from_D_to_Z}. By adding the acceptance criteria, the uniform sampling algorithm immediately transforms into a quantum weighted sampling algorithm in its own right. The thermalization properties of the algorithm can be improved by adding a ``problem'' term to the Hamiltonian.

Again inspired by QeMCMC, this extra Hamiltonian term (see equation \ref{eqn:H_prob_H_mix}) is designed to represent the BD action, so that the system can transition between states based on proximity in $S_{\epsilon}^{(d)}$ - a property required for efficient thermalization in low-temperature regimes. To do this, a qubit Hamiltonian representation of the BD action must be derived. We have shown in Appendix \ref{app:H_BD} that the following Hamiltonian approximates the action when $\epsilon\ll1$ while remaining cubic in complexity: 

\begin{equation}
\label{eqn:H_BD}
H_{BD_\varepsilon}^{(4)} =\frac{4}{\sqrt{6}} \sqrt\varepsilon\left[N- \varepsilon \sum_{k<m}^{N} C_{km} \Big(1-10\varepsilon \sum_{k<l<m} C_{kl}C_{lm}\Big) \right]  \,.
\end{equation}

This Hamiltonian for BD action could find use in a variety of quantum algorithms. We choose to add it to Equation \ref{eqn:H_TC_H_mix}, resulting in the complete Hamiltonian for BD-weighted sampling of causal sets:
\begin{equation}\label{eq:H_full}
\begin{aligned}
    H_{weighted} &= \gamma_{\text{constr}}  H_{\text{constr}} + 
    \gamma_{\text{prob}}  H_{\text{prob}} +
    \gamma_{\text{mix}} H_{\text{mix}} \\
    &= \gamma_{\text{TC}}  H_{\text{TC}} + 
    \gamma_{BD_\varepsilon}  H_{BD_\varepsilon} +
    \gamma_{\text{mix}} H_{\text{mix}}\,.
\end{aligned}
\end{equation}
Each contribution has a $\gamma$ coefficient that controls their relative weight. Of course, optimal choice of these algorithmic parameters will be crucial to an efficient quantum algorithm. As before, we normalize the Hamiltonian terms - for details see Appendix \ref{app:alphas}.

The full procedure is very similar to the uniform sampling algorithm. The Markov chain begins with a known causal set and updates by proposing a new configuration via a single measurement result after quantum real time evolution under Hamiltonian $H_{weighted}$ from equation \ref{eq:H_full}. The measured configuration is accepted if it is both a valid causal set (by calculating the value of $H_{TC}$ $\mathcal{O}(N^3)$ in classical complexity) and meets the Metropolis-Hastings criterion.

\textit{Simulated Results}\textemdash The proposed algorithms rely on the quantum scaling advantage of the QeMCMC. Proving the scaling (in terms of thermalization) of any MCMC procedure is difficult, and there is no analytical proof that QeMCMC outperforms classical methods \cite{orfi2024}. The algorithm relies on strong empirical evidence of improved thermalization. Due to the quadratic number of qubits required, simulations are limited to very small cardinality; so the empirics that we reproduce here are ineluctably very small. 

The absolute spectral gap ($\delta$) is employed as a figure of merit to quantify the convergence of a given Markov chain. It is particularly useful, as $\delta^{-1}$ bounds the thermalization time ($\tau_\alpha$) on both sides:
\begin{equation}
\left(\delta^{-1}-1\right) \ln \left(\frac{1}{2 \alpha}\right) \leq \tau_{\alpha} \leq \delta^{-1} \ln \left(\frac{1}{\alpha \min_C \nu(C)}\right) \ ,
\label{eqn:spec_therm}
\end{equation}
where $\alpha$ is the error in total variational distance \cite{levin2017}, and $\nu(C)$ is the (Boltzmann) probability of a given causal set C. The spectral gap can be calculated by finding the second largest eigenvalue of the transition matrix. To find this, one needs to know the probability of proposal from every causal set to every other causal set, as well as the probability of each possible move being accepted. Due to this, calculating the spectral gap is exponentially more expensive than an actual MCMC algorithm.

In the case of the uniform sampling algorithm, the probability of each move being accepted is one (in the transition probability matrix we only consider causal sets) meaning that proposals to non-causal sets are treated as non-proposals, adding to the diagonal. The probability of transition is dependent only on the proposal method used, meaning we can compare the different methods as fairly as possible using the spectral gap.

\begin{figure}
    \centering
    \includegraphics[width=0.9\linewidth]{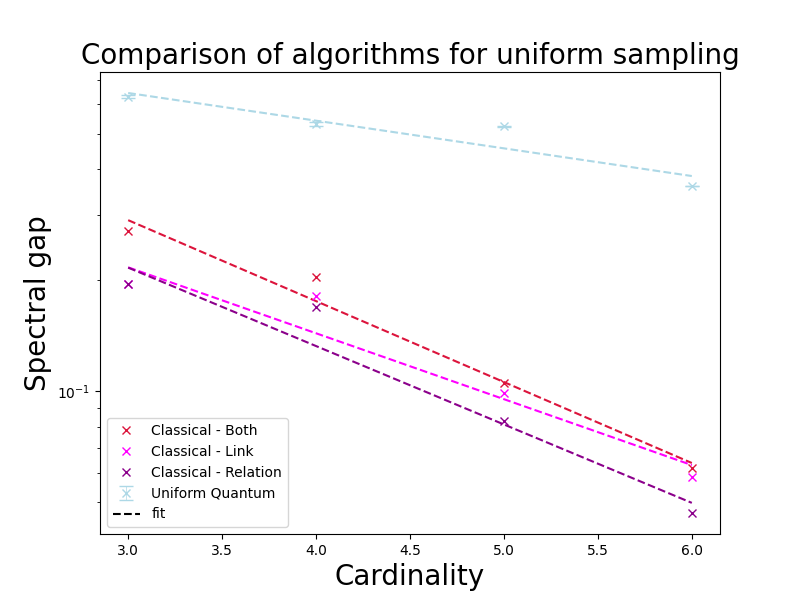}
    \caption{Spectral gap against cardinality for simulation of the quantum uniform sampling algorithm (blue) and classical strategies (shades of red). Each dotted line is an exponential decay fit, $\delta \propto e^{kN}$, with the exponent quoted for each: $k_Q = 0.17(5)$, $k_C =0.51(4)$, $k_{CL} =0.41(7)$, $k_{CR} =0.49(7)$.}
    \label{fig:uniform_spectral_gaps}
\end{figure}

In Figure \ref{fig:uniform_spectral_gaps}, the quantum uniform sampling algorithm is compared with the state of the art classical uniform sampling algorithm (split up into its two constituent ergodic moves) \cite{henson2015}. The spectral gap of the classical algorithm appears to show exponential decay, suggesting the exponential thermalization time that was observed in \cite{henson2006}. However, the quantum approach shows much slower exponential decay. Each set of results was fit to $\delta \propto e^{-kN}$, with the quantum proposal decaying cubically slower ($k_Q = 0.17(5)$) than the classical approach used in the literature ($k_C =0.51(4)$). Here, as in previous literature on QeMCMC, ``cubic'' scaling advantage refers the reduction of $k$ by a factor of $3$.

\begin{figure}
    \centering
    \includegraphics[width=0.9\linewidth]{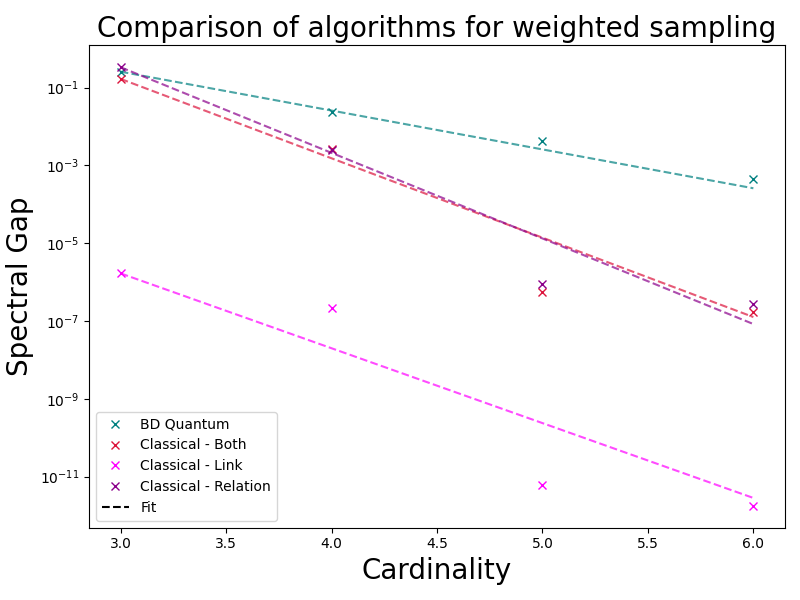}
    \caption{Spectral gap against cardinality for simulation of the quantum BD sampling algorithm (green) and classical strategies (shades of red). Temperature, $T = 0.004$. Each dotted line is an exponential decay fit, $\delta \propto e^{-kN}$, with the exponent quoted for each: $k_Q = 2.09(8)$, $k_C = 4.3(4)$, $k_{CL} =4(2)$, $k_{CR} =5.1(2)$.}
    \label{fig:BD_sampling_spectral_gaps}
\end{figure}

For the more complex problem of sampling from the causal set partition function, we employ our weighted quantum sampling algorithm and compare it against the classical proposal (with Metropolis-Hastings acceptance) in Figure \ref{fig:BD_sampling_spectral_gaps}. In this case, the quantum method performs considerably better than the classical approaches; however there are variables to consider. The difficulty of the problem is inextricably linked to the temperature, $1/\beta$. For the spin-glass problems explored in QeMCMC literature so far, low temperatures make the sampling problem harder, and in high temperatures the problem converges to uniform sampling - which is comparatively trivial. The temperature dependence is less obvious when we do not restrict the sample space in the same way. The temperature dependence is explored in Appendix \ref{app:temp_analysis}. As the temperature dependence is very extreme for the classical algorithm, we do not explicitly quantify a quantum scaling advantage. In Figure \ref{fig:BD_sampling_spectral_gaps}, we choose a relatively low temperature regime ($T = 0.004$), yet before the extrema where it simply becomes the optimization problem of finding the causal set that corresponds to the lowest action. See Appendix \ref{appendix_F} for a discussion on the physical meaning of the temperature in our case. Dimension is another variable here, specifically the dimension of the BD action that you choose to take. It appears that for 2d BD action, the problem is simpler compared to higher dimensions; so we have focused on the harder 4d case. We leave the study of the algorithm in higher dimensions to further work.
% this is in a dimensionally restricted setting, where the sample space is deliberately restricted to manifold-like causal sets. 

\textit{Conclusion}\textemdash In this letter, we have proposed two quantum algorithms that bridge the gap between quantum computing and fundamental physics. The first is capable of uniformly sampling from the space of causal sets more efficiently than known classical algorithms. The second algorithm performs weighted sampling according to the Benincasa-Dowker action, exploring the causal set partition function. Both algorithms exhibit a super-quadratic scaling, and inherit the noise-resilience of their parent algorithm, QeMCMC \cite{layden2023}. Along the way, novel contribution to quantum computing has been made by constraining the QeMCMC, while a qubit Hamiltonian was necessarily derived, contributing to causal set theory. 

We achieve a cubic quantum scaling advantage in the uniform sampling algorithm and we do not to explicitly quantify the weighted sampling algorithm, due to its temperature dependence. We also sample from a range of $\gamma$ parameters; optimization of these is very likely to result in better scaling. One of the limitation of our approach is the requirement of a quadratic number of qubits. This results in a severe limitation when simulating the algorithm either by spectral gap or directly. Future work should focus on numerics for larger systems, which should include exploiting coarse-graining of the system to access larger cardinality \cite{ferguson2025}. It would be very interesting to scale the simulations by using existing hardware, both digital or even analogue devices to investigate the effects of noise on the algorithm \cite{arai2025}. Classical simulation through a quantum inspired approach would also be an interesting direction to explore \cite{christmann2024}. The current state of quantum hardware means that we cannot yet implement the algorithm to truly investigate causal set dynamics. However, in the near future, this quantum algorithm could explore causal set dynamics for larger $N$, and in more extreme conditions, than is classically possible. This would open the door to studying causal-set dynamics in non-dimensionally restricted settings which is currently under-explored.

\textit{Acknowledgments}\textemdash
We would like to thank Fay Dowker and Sumati Surya for useful discussions.
P.W. and A.N. acknowledge funding from STFC with grant number ST/W006537/1, S.F. support
by EPSRC DTP studentship grant EP/W524311/1 and P.W. funding from EPSRC grants EP/T001062/1 and EP/Z53318X/1. A.N. is funded by the President’s PhD Scholarship from Imperial College London

%\petros{Arad and Stuart: do we want to thank anyone else? Either peson we discussed with or any other funding?}\arad{I added my imperial funding}

\bibliography{References.bib}

\appendix

\section{Action and Hamiltonian in arbitrary dimension} \label{app:H_BD}
\label{appendix_arbitrary_d}
Unlike the Einstein-Hilbert action, the BD action is dimension-dependent. That is, in order to recover the Einstein-Hilbert action in the continuum limit, one has to sum over the abundances $N_j$ with different coefficients in different dimensions. The expression for the BD action in an arbitrary spacetime dimension d is given by \cite{dowker2013causal}:
\begin{equation}
    \begin{aligned}
\label{eqn:smeared_BD_action_in_d}
\centering
    S_{\varepsilon}^{(d)} & =-\alpha_d\Big(\frac{\ell_c}{\ell_p}\Big)^{d-2} \varepsilon^{2/d}\left[N+\frac{\beta_d}{\alpha_d} \varepsilon \sum_{j=0}^{N-2}f_d(j, \varepsilon) N_j \right] \\
f_d(j, \varepsilon) & =(1-\varepsilon)^j\sum_{k=1}^{\lfloor\frac{d}{2}\rfloor+2}C_k^{(d)}{j\choose k-1}\Big(\frac{\varepsilon}{1-\varepsilon}\Big)^{k-1},
\end{aligned}
\end{equation}
where $\alpha_d,\ \beta_d$, and $C_k^{(d)}$ are dimension-dependent constants, $\ell_p$ is the Planck length defined by $\ell_p^{d-2}=8\pi G$, and $\ell_c$ is the discreteness length which is supposedly at the same order of magnitude as the Planck length. Assuming $\ell_c=\ell_p$, the action for $d=4$ is given be Equation~\eqref{eqn:smeared_BD_action}.

Up to first order in $\varepsilon$, $f_d$ is given by
\begin{equation}
    f_d(j,\varepsilon)\approx(1-j\varepsilon)(1+C_2^{(d)}j\varepsilon)\approx 1+(C_2^{(d)}-1)j\varepsilon+\mathcal{O}(\varepsilon^2).
\end{equation}
For reference, we can write down the explicit form of $C_2^{(d)}$ using the general expression in \cite{glaser2014closed}:
\begin{equation} 
C_2^{(d)} \sim
\left\{
\begin{aligned}
 &1-{d+1\choose \frac{d}{2}},\ \ \  \  \text{even d}\\
&1-\frac{(2d+1)!!}{(d+1)!},\ \ \text{odd d}
\end{aligned}
\right.
\end{equation}

Following the same steps as Appendix \ref{app:H_BD}, we find the following Hamiltonian that approximates the BD action for $\epsilon\ll1$ in dimension d
\begin{equation}
\begin{aligned}
 H_{BD_\varepsilon}^{(d)} &=-\alpha_d\Big(\frac{\ell_c}{\ell_p}\Big)^{d-2} \varepsilon^{2/d}\Bigg[N +\\  &\frac{\beta_d}{\alpha_d} \varepsilon \sum_{k<m}^{N} C_{km} \Big(1+(C_2^{(d)}-1)\varepsilon \Lambda_{km}\Big) \Bigg] \,.
\end{aligned}
\end{equation}
This too has at most cubic interactions in the causal matrix elements.

\section{Hamiltonian representation of the original BD action}
The quantum algorithm so far developed only works in the regime $\epsilon\ll1$. The original proposal for the BD action in \cite{benincasa2010}, however, uses $\epsilon=1$ as it gives a very simple and tractable expression for the action. Here we show that it is possible to write down a Hamiltonian with at most cubic interactions that equates the BD action, but with the trade-off of increasing the number of qubits. 

In 2d, for $\epsilon=1$, the BD action simplifies to
\begin{equation}
    S_{BD}=2(N-2N_1+4N_2-2N_3).
\end{equation}

The first step is to write this action in terms of the causal matrix since its elements are the qubits. Define
\begin{equation}
\label{def_M}
    M_{km}^{(j)}=\delta(\Lambda_{km},j),
\end{equation}
where $\delta$ is the Kronecker delta. $M_{km}^{(j)}$ is a binary variable that is equal to 1 if and only if there are $j$ elements between $i$ and $j$, excluding $i$ and $j$. This new matrix allows us to write a $BD$ Hamiltonian that is exactly equal to the above BD action
\begin{equation}
    H_{BD}=2N-2\sum_{k<m}\big(C_{km}M_{km}^{(0)}-2M_{km}^{(1)}+M_{km}^{(2)}\big).
\end{equation}
So if we introduce $M_{km}^{(j)}$ for $1\leq k<m\leq N$ and $0\leq j\leq2$ as independent qubits, the BD Hamiltonian would consist of linear and quadratic terms. So far this adds $\mathcal{O}(N^2)$ new qubits, but then we should add Hamiltonian constraints to ensure that $M_{km}^{(j)}$ and $C_{km}$ are consistent. In other words, we need penalty terms in the total Hamiltonian to implement the Kronecker delta in~\eqref{def_M}.

To do so, we first introduce $\mathcal{O}(N^3)$ ancilla qubits $P_{k\ell m}$ which are meant to equate the product $C_{k\ell}C_{\ell m}$. This will give us the following linear expression:
\begin{equation}
    \Lambda_{km}=\sum_{\ell=k+1}^{m-1}P_{k\ell m}.
\end{equation}
To ensure $P_{k\ell m}=C_{k\ell}C_{\ell m}$, we need a penalty term of the form $\lambda\left(P_{k\ell m}-C_{k\ell}C_{\ell m}\right)^2$. As these are binary variables, we add the following cubic penalty terms to the Hamiltonian
\begin{equation}
    H_{k\ell m}=\lambda \left(P_{k\ell m}-2P_{k\ell m}C_{k\ell}C_{\ell m}+C_{k\ell}C_{\ell m}\right).
\end{equation}
%(\arad{To Petros: Is it valid not to worry about operator ordering?}) 

Now, to enforce a Kronecker delta on $\Lambda_{km}$, we need ancilla qubits $s_{km}^i$ that are meant to represent the binary representation of $\Lambda_{km}$: 
\begin{equation}
    \sum_{\ell=k+1}^{m-1}P_{k\ell m}= s_{km}^{0}2^0+s_{km}^12^1+...+s_{km}^n2^n,
\end{equation}
where $n+1$ is the number of digits in the binary representation of $\Lambda_{km}$. Again, this equality has to be enforced using penalty terms in the Hamiltonian. We claim that this can be done using cubic penalty terms and $\mathcal{O}(N\log N)$ ancilla qubits for each $k$ and $m$, making a total of $\mathcal{O}(N^3\log N)$ ancilla qubits.

To see this, assume that given the qubits $a_0,...,a_N$, we have already enforced the following binary representation using ancilla qubits
\begin{equation}
    a_1+...+a_{N-1}=b_02^0+b_12^1+...+b_n2^n.
\end{equation}
To add a new qubit $a_N$ to the above sum, first introduce ancilla qubits $c_0,...,c_{n+1}$, and $d_0,...,d_n$, and for each $0\leq i\leq n+1$. If we could enforce $c_i=b_i\oplus d_{i-1}$ and $d_i=b_id_{i-1}$, then we would have the following representation 
\begin{equation}
    a_1+...+a_{N}=c_02^0+c_12^1+...+c_{n+1}2^{n+1}.
\end{equation} 
We can ensure that $c_i=b_i\oplus d_{i-1}$ and $d_i=b_id_{i-1}$ hold if we add the following two terms to the Hamiltonian
\begin{align}
    &H_{c_i}=\lambda(c_i-b_i-d_{i-1}+2b_id_{i-1})^2\\\nonumber    &=\lambda(c_i+b_i+d_{i-1}-2c_ib_i-2c_id_{i-1}-2b_id_{i-1}+4c_ib_id_{i-1}),\\
    &H_{d_i}=\lambda(d_i-b_id_{i-1})^2\\\nonumber
&=\lambda(d_i+b_id_{i-1}-2d_ib_id_{i-1}).
\end{align}
In the above, we use $d_{-1}\equiv a_N$. 

Note that if $\log_2N$ was not an integer, then we did not need $c_{n+1}$. For adding the $N$th term of the summation $a_N$, we needed $\mathcal{O}(\log N)$ ancilla qubits and cubic interaction terms, which proves our earlier claim.

Let us get back to $s_{km}^i$. Let $j_i$ be the $i$th digit of $j$ in binary representation. Equation \eqref{def_M} can be rewritten as 
\begin{equation}
    M_{km}^{(j)}=\prod_{i=0}^n\left[1-(j_i-s^i_{km})^2\right].
\end{equation}

Equating the two sides using a penalty term would be difficult because the right side is a high-degree polynomial. But since $j_i$ and $s^i_{km}$ are all binary, the right-hand side can be reconstructed from cubic polynomial using $n\sim\mathcal{O}(\log N)$ ancilla qubits for each $k,m$.

Therefore, the BD action for $\epsilon=1$ can be used in QeMCMC using a cubic Hamiltonian and a total of $\mathcal{O}(N^3\log N)$ qubits.

\section{From decoherence functional to partition function}
\label{appendix_from_D_to_Z}

Understanding the dynamics of a causal set is a challenging problem. An important development in this direction was the construction of a class of classical sequential growth models. In these models, the causal sets grow one element at a time, and each transition has a certain probability that is determined based on the requirement of discrete general covariance and Bell causality \cite{rideout1999classical}. However, in \cite{rideout2001evidence,brightwell2010continuum} it was shown that the continuum limit of a classical sequential model is given by an almost-semiorder, and hence is not manifoldlike. Therefore, recovering the classical Lorentzian spacetime manifolds requires models that go beyond classically stochastic growth models.

Classical sequential growth models belong to the level zero of the hierarchy of measure theories defined in \cite{sorkin1994quantum,sorkin1995quantum}. In this sense, the next natural step for developing dynamics of causal sets is to define a consistent decoherence functional for causal sets to have a level one measure theory. This would provide a theory of quantum gravity for causal sets. Taking the decoherence functional as the fundamental object of the theory, one can then proceed to define a histories Hilbert space, as outlined in \cite{dowker2010hilbert}. 

As a simple example for a non-relativistic particle in $d$ spacetime dimensions, the decoherence functional between two trajectories $\gamma_1$ and $\gamma_2$ is given by
\begin{align}
\label{non-rel_decoherence_functional}
    D\big(\Bar{\gamma},\gamma\big)=e^{-iS[\Bar{\gamma}]+iS[\gamma]}\ \delta^{d-1}_{\gamma(T),\Bar{\gamma}(T)}\ \psi^*\Big(\Bar{\gamma}(0)\Big)\psi\Big(\gamma(0)\Big)\,,
\end{align}
where $\delta^{d-1}$ is the $d-1$ dimensional Dirac delta function, and $\psi$ is the initial wavefunction of the particle. The action encodes the dynamics, the delta function enforced on a final time $T$ hypersurface ensures that only the trajectories that end at the same point can interfere, and $\psi$ is the initial wave function. Now, in a level one theory in the hierarchy of measure theories in the sense of \cite{sorkin1994quantum}, one would start with the above decoherence functional as the fundamental object and build a histories Hilbert space. One can show that this is isomorphic to the ordinary Hilbert space of square-integrable functions for simple cases like a particle in a simple harmonic oscillator \cite{dowker2010hilbert}. 

By analogy with \eqref{non-rel_decoherence_functional}, one is tempted to define a decoherence functional for $N$-element causal sets as
\begin{align}
    D\big(\{\Bar{C}\},\{C\}\big)=e^{-iS[\Bar{C}]+iS[C]}\ \Theta(\Bar{C},C).
\end{align}
Here $\Theta$ is the interference kernel that determines what causal sets as full spacetime histories are allowed to interfere. Due to the non-local structure of causal sets and the difficulty to properly define spatial hypersurfaces in causal sets, it is not straightforward to find what $\Theta$ should be. Yet it will play a significant role in determining the Hilbert space of N-element causal sets. In particular, if $\Theta$, apart from the initial wave function factors, turns out to depend solely on some "late time" properties of $C$ and $\Bar{C}$, then the histories Hilbert space will be at most of the size of the space spanned by all such late-time states. Also note that a Hilbert space will be defined for N-element causal sets, and since the number of these causal sets grows as $2^{N^2/4}$, the size of the causal set Hilbert space will most probably grow with time \footnote{Here $N$ (or the total volume of the spacetime as its continuum counterpart) serves as the natural time parameter.}. Nevertheless, unitarity is still present in the sense of preservation of the inner product of states under the dynamics given that D would be a consistent decoherence function:
\begin{align}
    D(\Bar{C},C)=\sum_{\substack{|\Bar{C}'|=|C'|=N+1\\ \Bar{C}'\big|_N=\Bar{C},\ C'\big|_N=C}}D(\Bar{C}',C')\,.
\end{align}
This unitarity condition puts a strong constraint on the choice of $\Theta$.

Assuming we have a choice of $\Theta$ that preserves the unitarity condition, one may choose to write a single path integral formalism (instead of the above double path integral) for causal sets by integrating out $\Bar{C}$. 
\begin{align}
    \frac{\mu(C)e^{iS[C]}}{Z[N]}=\sum_{|\Bar{C}|=N}D(\Bar{C},C).
\end{align}

Note that in the above definition, we have separated the $C$ dependence and $N$ dependence of the sum $\sum_{\Bar{C}}e^{-iS[\Bar{C}]}\ \Theta(\Bar{C},C)$ into a measure factor $\mu$ and a partition function
\begin{align}
    Z[N]=\sum_{|C|=N}\mu(C)e^{iS[C]}.
\end{align}
Therefore, the choice of the interference kernel $\Theta$ manifests itself in the measure factor inside the path integral. In general, apart from an initial wave function factor, $\mu(C)$ also has a dependence on the bulk of $C$.

In the studies on the above path integral, it is typically assumed that $\mu(C)$ can at most depend on an initial wave function factor and does not depend on the causal structure of $C$. Therefore, although it seems natural to sum over all causal sets just with the amplitude $\exp(iS)$, this puts a non-trivial constraint on the interference kernel. Also note that the decoherence functional provides a more fundamental starting point compared to the partition function. It provides a well-defined histories Hilbert space for quantum causal sets. However, starting from the partition function, it is far from obvious what the Hilbert space is and what we are actually computing when we insert functions in the path sum or restrict it to a certain class of causal sets.

Nevertheless, as long as we are interested in the problem of sampling from causal sets and finding suitable quantum algorithms, we can work with the equal-weight partition function

\begin{align}
\label{Z_unif}
    Z_{\rm unif}[N]=\sum_{|C|=N}e^{iS[C]}.
\end{align}

For sampling from the correct partition function $Z[N]$, we would then have to do importance sampling with weights $\mu(C)$, or use $\mu$ in our Metropolis-Hastings with additional factors of $\mu(C')/\mu(C)$. With this understanding, we remove the $\rm unif$ subscript and treat \eqref{Z_unif} as the main partition function.

\section{Normalisation of Hamiltonian terms}

\label{app:alphas}
The scale of the $H_{mix}, H_{TC}$ and $H_{BD}$ varies between cardinalities, meaning that normalising their weight is convienient. This means that we can consistently use the relative $\gamma$ weights (which we define to sum to $1$) to control the magnitude of contribution from each term. For BD action we simply follow the procedure from \cite{layden2023}, ignoring higher order terms in $\epsilon$ and trivial terms in $H_BD$.

\begin{equation}
    \alpha_{BD} = \frac{||H_{mix}||_{\mathrm{F}}}{||H_{BD}||_{\mathrm{F}}} \approx  \frac{\sqrt{q}}{\sqrt{4\epsilon ^4 \sum_{k<m}^q 1  
    }} = \frac{\sqrt{q}}{\sqrt{4\epsilon^4q}} \,,
\end{equation}

where $\|M\|_{\mathrm{F}}=\operatorname{tr}\left(M^{\dagger} M\right)^{1 / 2}$ is the Frobenius norm\footnote{In the case of a Hamiltonian expressed as Pauli strings, this simplifies to the sum of squares of the coefficients of the strings.} of a matrix $M$ and we use $q = \frac{N(N-1)}{2}$ to denote the number of qubits in the hamiltonian. 

To compute $\alpha_{TC}$ requires more care. The size of the energy gap between causal sets and non-causal sets cannot decay for large cardinality, meaning we should not normalize using the Frobenius norm as doing so would keep the maximum value of $H_{TC}$ approximately constant. Instead we must normalize with respect to the first excited state, $H_{TC_1}$.

\begin{equation}
     \alpha_{TC} = \frac{||H_{mix}||}{||H_{TC_1}||} =  \frac{\sqrt{q}}{\frac{8}{{N\choose 3}}}\,,
\end{equation}

where the denominator normalizes the sum over 3-pairs of elements in $H_{TC}$ so the first excited state energy, $H_{TC_1} = 1$. Qualitatively, this means that we have large degenerate energy levels, algorithmically this adds to the computational complexity of the Hamiltonian simulation. The scaling of such algorithms are generally dependent on either the spectral norm or the L$_1$-norm of the Hamiltonian \cite{low2019}, meaning that the Hamiltonian simulation component of the quantum algorithm scales with the norm of $H_{TC}$.

\section{Temperature dependence of weighted sampling} \label{app:temp_analysis}

As with any MCMC algorithm of this type, there is a strong dependence on temperature when weighted sampling according to BD action. The QeMCMC has been shown to work especially well for Ising models in the low-temperature setting, where the energy landscape is particularly ``glassy''. In our case, the relationship with temperature is less clear. Classical literature \cite{cunningham2020} has reported slow thermalization for extreme $\beta$, so in this section we discuss the efficacy of the quantum algorithm in such conditions.

\begin{figure}
    \centering
    \includegraphics[width=1\linewidth]{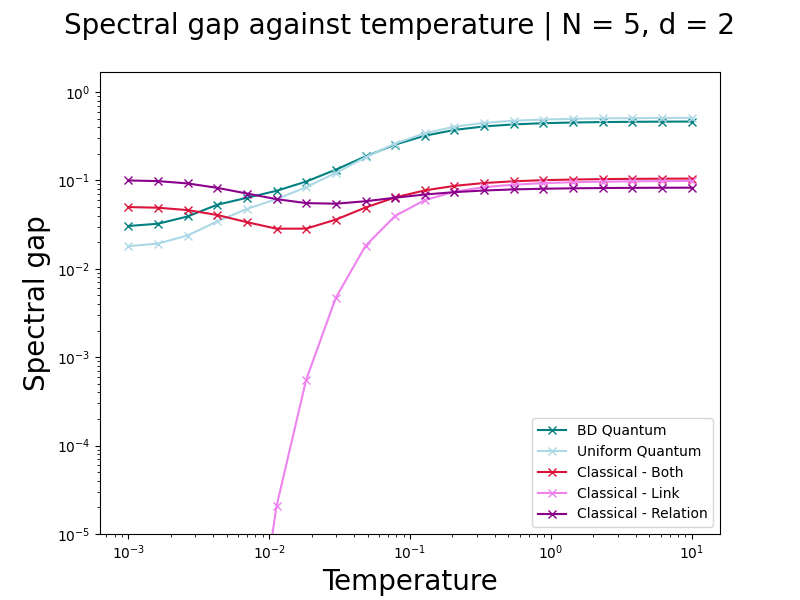}
    \caption{Temperature dependence of spectral gap for each proposal method, uniform quantum (light blue), BD quantum (green) and classical
strategies (shades of red) for 4d BD action. Cardinality, $N =5$ and dimension, $d = 2$.}
    \label{fig:temp_gaps_N5_d2}
\end{figure}

Figure \ref{fig:temp_gaps_N5_d2} displays the relationship between spectral gap and temperature for both quantum and classical algorithms at $N=5$ for 2d BD action. The classical relation move seems to perform very well across temperatures, while the link move performs poorly in low temperature regimes. This is likely due to the fact that in \textit{2d}, the BD action results in a much less "glassy" cost function landscape, so the quantum method is not required. In fact, in this case it appears that the low-temperature case is easier than high temperature - meaning the problem of finding the lowest action may be relatively trivial.

\begin{figure}
    \centering
    \includegraphics[width=1\linewidth]{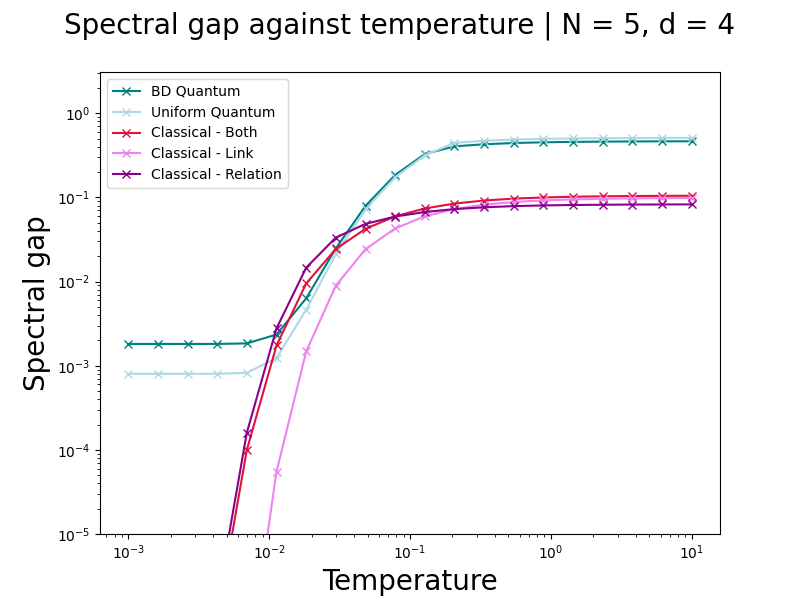}
    \caption{Temperature dependence of spectral gap for each proposal method, uniform quantum (light blue), BD quantum (green) and classical
strategies (shades of red) for 2d BD action.  Cardinality, $N =5$ and dimension, $d = 4$.}
    \label{fig:temp_gaps_N5_d4}
\end{figure}

In the 4d case (Figure \ref{fig:temp_gaps_N5_d4}) however, both classical algorithms fail at low temperatures, where as the simulated quantum method does not exhibit this critical slowing down. This is a strong indication that the quantum algorithm works, while the increased spectral gap between the quantum algorithm with the BD Hamiltonian term is an indication that the BD Hamiltonian is working as expected. If one were to take a lower temperature and re-plot Figure \ref{fig:BD_sampling_spectral_gaps} from the main text, the quantum scaling advantage could be arbitrarily high, hence why we do not attempt to quantify it.

\section{Why do we sample from partition function?}
\label{appendix_F}
An important question that the partition function has to answer is whether it suppresses non-manifoldlike causal sets. Such causal sets dominate the space of partial orders. A class of 3-layered causal sets, called KR orders, entropically dominate the set of all causal sets and they are followed by bi-layer orders and $k$-layer orders for $k\geq 3$ \cite{kleitman1975asymptotic}. In a series of analytic studies \cite{loomis2017suppression,mathur2020entropy,carlip2023path,carlip2024einstein}, it was shown that the restricted partition function over KR orders (and in fact all k-layer orders with $k\ll N$) is exponentially suppressed. This is an important hint that the path integral dynamics might naturally favor manifoldlike causal sets while suppressing non-manifoldlike causal sets.

These analytic results rely on the cancellation of oscillations in $\exp({iS_\varepsilon^{(d)}[C]})$. However, numerical studies are more tractable in the Euclidean setting. So one would introduce an auxiliary parameter $\beta$ in $\exp({i\beta S_\varepsilon^{(d)}[C]})$. Then the amplitude is analytically continued to an actual statistical distribution using $\beta\rightarrow i\beta$. This enables MCMC sampling of causal sets. Note that in 4d, instead of assuming $\ell_c=\ell_p$ and defining a new auxiliary parameter, we can identify $\beta=\left(\frac{\ell_c}{\ell_p}\right)^2$ and analytically continue it. In this sense, low temperatures correspond to a larger discreteness scale, and high temperatures correspond to a smaller discreteness, closer to the continuum limit.

\section{Implementation details}

This section provides details of the simulated constrained QeMCMC implementations used to create the figures in the main text. The full python code can be found on github \cite{Ferguson_SamplingCausalSets}.

When implementing time evolution under a Hamiltonian on quantum hardware, it is necessary to first translate the binary variables to spins, $s$. This is straightforward by making the substitution: $C_{ij} = \frac{1-s_{ij}}{2}$ before replacing the spin variables by the Pauli $Z$ matrix.

The transitive closure constraint (equation \ref{eqn:H_TC} in the main text), ignoring trivial constant terms becomes

\begin{equation}
\label{eqn:H_pop}
\begin{aligned}
     H_{TC} =&   \sum_{i<j<k} \frac{1}{8}(-Z_{ij}-Z_{jk}+Z_{ik}+\\
    & Z_{ij}Z_{jk}-Z_{ij}Z_{ik}-Z_{jk}Z_{ik} +Z_{ij}Z_{jk}Z_{ik})\,.\\
\end{aligned}
\end{equation}

Similarly, for the BD term (equation \ref{eqn:H_BD} in the main text)
\begin{equation}\label{eq:smeared_BD_ham}
\begin{aligned}
H_{BD\varepsilon} =& [\frac{4}{\sqrt{6}} \sqrt{\varepsilon} N] - [\frac{2}{\sqrt{6}}\varepsilon^\frac{3}{2} \sum_{k<m}^{N} 1- Z_{km}] + \\
& [\frac{5}{6}\varepsilon^\frac{5}{2} \sum_{k<l<m}^{N} 1 - Z_{km}-Z_{kl}-Z_{lm}+\\
&Z_{kl}Z_{lm}+Z_{kl}Z_{km}+Z_{lm}Z_{km} -Z_{kl}Z_{lm}Z_{km})] \,.
\end{aligned}
\end{equation}

The time evolution was simulated by first order trotter decomposition \cite{suzuki1990}, with quantum circuits simulated by qulacs simulator \cite{suzuki2021qulacs}.

In all simulated experiments, $H_{mix} = \sum_iX_i$. Each $\gamma$ is sampled from a range, to avoid ``over-fitting'' the parameters. To simplify things, we enforced $\gamma_{mix}+\gamma_{TC}+\gamma_{BD_\epsilon} = 1$ and sampled the two variables $r_{TC}$ and $r_{BD}$ where $\gamma_{TC} = r_{TC}$, $\gamma_{BD} = (1-r_{TC})r_{BD}$, $\gamma_{BD} = (1-r_{TC})(1-r_{BD})$. In the uniform sampling experiments, each was sampled from ranges: $r_{TC} = \{0.7,0.9\}$, $r_{BD} = \{0,0\}$. For weighted sampling the following ranges were used: $r_{TC} = \{0.7,0.9\}$, $r_{BD} = \{0.05,0.02\}$. These are not optimal, and there is large scope for improvement here. The evolution time (number of trotter steps) was sampled between $t = \{3,10\}$. The BD action smearing term was taken to be $\epsilon = 0.1$, which is actually relatively large compared to some estimations by Sorkin suggesting that it should be $\approx 10^{-20}$ \cite{sorkin2007}.

Note that for each cardinality, there exists only one space of all causal sets. This means that no error has been reported for the classical approaches in Figures \ref{fig:uniform_spectral_gaps} and \ref{fig:BD_sampling_spectral_gaps} in the main text as the spectral gap methods are exact. For the quantum algorithms however, as we sample the parameters, we can quantify the error. For each causal set, one can use the state-vector produced by a particular $\gamma$ and $t$ as the probability of proposal to other causal sets. Combining these samples, the transition matrix and thus spectral gap can be calculated, while the error in spectral gap can be calculated by the jackknife method from the sampled state-vectors. In this case, only 10 jackknife iterations were sufficient to minimize the errors sufficiently.

%\begin{equation}
%\label{eqn:H_prob_H_mix}
%    H = (1-\gamma)H_{prob} + \gamma H_{mix} + %PH_{constraint}
%\end{equation}

\end{document}